\documentclass[12pt,times,draftclsnofoot,a4paper]{elsarticle}

\usepackage{graphicx}
\usepackage{lineno}
\usepackage{xspace}
\usepackage{lipsum}
\usepackage{stackengine}
\usepackage{titlesec}
\usepackage{xcolor}
\usepackage{comment}
\usepackage[bitstream-charter]{mathdesign}
\usepackage[T1]{fontenc}
\usepackage{geometry, float}
\usepackage[normalem]{ulem}
\usepackage{stackengine}
\usepackage{siunitx}
\usepackage{hyperref}
\hypersetup{colorlinks=true}

\usepackage{cleveref}

\journal{Nuclear Instruments and Methods in Physics Research A}

\begin{document}

\begin{frontmatter}

\setcounter{page}{0}
\title{Beam test of a Pb/SciFi prototype for the Barrel Imaging Calorimeter at the Electron--Ion Collider}
\author[skku]{Hyungjun~Lee\fnref{fn1}}\ead{leehy@cern.ch}
\author[knu]{Changhui~Lee\fnref{fn1}}\ead{logan12@knu.ac.kr}
\author[pnu]{Jaehyeok~Ryu\fnref{fn1}}\ead{jaehyeok.ryu@cern.ch}

\author[kangwon]{Geunpil~An}
\author[skku]{Joonsuk~Bae}
\author[skku]{Yunseul~Bae}
\author[pnu]{Jeongsu~Bok\corref{cor1}}\ead{jeongsu.bok@cern.ch}
\author[yonsei]{Yun~Eo}
\author[skku]{Wooseok~Ham}
\author[pnu]{Yoonha~Hong}
\author[anl]{Manoj~Jadhav}
\author[yonsei]{Seo~Yun~Jang}
\author[kangwon]{Jinryong~Jeong}
\author[knu]{Hyon-Suk~Jo}
\author[anl]{Sylvester~Joosten}
\author[skku]{Beomkyu~Kim}
\author[anl]{Bobae~Kim}
\author[pnu]{Chong~Kim}
\author[skku]{Dongguk~Kim}
\author[kangwon]{Minsuk~Kim}
\author[knu]{Shin~Hyung~Kim}
\author[knu]{Wonjun~Ko}
\author[knu]{Sehwook~Lee}
\author[yonsei]{Sanghoon~Lim}
\author[anl]{Jessica~Metcalfe}
\author[regina]{Zisis~Papandreou}
\author[skku]{Sangwoo~Park}
\author[knu]{Junseop~Shin}
\author[yonsei]{Hwidong~Yoo}
\author[anl]{Maria \.{Z}urek}

\cortext[cor1]{Corresponding author at: Department of Physics, Pusan National University, Busan, 46241, Republic of Korea.}
\fntext[fn1]{These authors contributed equally to this work as co-first authors.}

\address[skku]{Department of Physics, Sungkyunkwan University, Suwon 16419, Republic of Korea}
\address[knu]{Department of Physics, Kyungpook National University, Daegu 41566, Republic of Korea}
\address[pnu]{Department of Physics, Pusan National University, Busan 46241, Republic of Korea}
\address[kangwon]{Department of Physics, Kangwon National University, Gangneung 25457, Republic of Korea}
\address[yonsei]{Department of Physics, Yonsei University, Seoul 03722, Republic of Korea}
\address[anl]{Argonne National Laboratory, Lemont, IL 60439, U.S.A.}
\address[regina]{Department of Physics, University of Regina, Regina, Saskatchewan, S4S 0A2, Canada}

\begin{abstract}
A Lead-Scintillating Fiber (Pb/SciFi) prototype for the Barrel Imaging Calorimeter (BIC) at the Electron--Ion Collider (EIC) was tested with electron beams at the CERN PS T10 beam line in August 2024.
The prototype consisted of unit modules with a sampling structure of lead sheets and scintillating fibers, corresponding to a total depth of approximately $10.9\,X_{0}$.
Beam tests were performed with electron momenta between 0.5 and 3~GeV/$c$ to evaluate the energy and timing performance of the prototype.
This study characterizes the performance of a Pb/SciFi prototype and provides input for future beam tests, calibration and readout optimization, and the development of larger-scale prototypes.

\end{abstract}

\end{frontmatter}


\section{Introduction} \label{sec:intro}
The Electron-Ion Collider (EIC), to be constructed at Brookhaven National Laboratory, is a next-generation facility designed to explore the structure of nucleons and nuclei with unprecedented precision~\cite{AbdulKhalek:2021gbh}. By colliding polarized electrons with polarized protons and a wide range of ion species over a broad center-of-mass energy range, the EIC will enable systematic studies of partonic dynamics, including parton distribution functions, transverse-momentum-dependent distributions, and generalized parton distributions. These include the emergence of nucleon mass and spin, as well as the three-dimensional imaging of quarks and gluons inside the nucleon and nucleus. Achieving these physics goals requires a detector system with excellent tracking, particle identification, and calorimetric performance over a large acceptance. 

The Barrel Imaging Calorimeter (BIC) serves as the electromagnetic calorimeter in the barrel region of the ePIC detector, providing electron and photon measurements over the pseudorapidity range $-1.71 \lesssim\eta\lesssim 1.31$. 
To satisfy the physics goals, the BIC is required to have a moderate energy resolution of $10\%/\sqrt{E} \oplus 2-3\%$, provide electron identification and electron-pion separation over the energy range from 1 to 50~GeV, and reconstruct photons while distinguishing them from $\pi^0$ decays up to 10~GeV~\cite{EIC:Requirements}.
To meet these performance requirements, the BIC adopts a hybrid calorimeter concept consisting of a high-granularity imaging section followed by a longitudinally segmented bulk electromagnetic calorimeter.
The imaging section comprises several layers of AstroPix~\cite{Steinhebel:2025pra,Kim:2025yyh} high-voltage CMOS (HV-CMOS) monolithic active pixel sensors, interleaved with Pb/SciFi sampling layers providing high-granularity imaging of the electromagnetic shower development.
The Pb/SciFi bulk sector, based on the GlueX barrel electromagnetic calorimeter design~\cite{Leverington:2008zz,Beattie:2018xsk,GlueX:2020idb}, contributes substantially to the longitudinal shower containment and the total energy measurement, together with the imaging section.
The full calorimeter covers the barrel region with a longitudinal length of 435~cm along the beam axis and has an inner radius of 82.5~cm.
Its total depth in radiation length ($X_0$) corresponds to approximately 17.1~$X_0$ at $\eta=0$, with an effective radiation length of 1.45~cm.

As part of the BIC R\&D program, the Korean BIC group has been developing local fabrication and assembly capabilities for Pb/SciFi calorimeter modules.
Prototype modules produced in Korea were tested with electron beams at the CERN PS T10 facility to validate their performance. 

The remainder of this paper is organized as follows. The experimental setup, including the CERN PS T10 beam line, detector configuration, and Pb/SciFi prototype, is described in Section~\ref{sec:setup}. The beam alignment, channel equalization, and calibration procedures are presented in Section~\ref{sec:calib}. Section~\ref{sec:results} discusses the beam-test results on energy response, timing performance, and longitudinal shower development, and Section~\ref{sec:summary} concludes the paper.

\section{Experimental Setup}
\label{sec:setup}

\subsection{CERN PS Test Beam Facility}
We performed the experiment at the CERN PS T10 beam line. 
The T10 beam line at CERN's PS East Area provides secondary beams generated from 24 GeV/$c$ protons slow-extracted from the Proton Synchrotron~\cite{Bernhard:2021jeb}. 
The beam contains electrons, pions, and muons, with minor contributions from protons and kaons, and its composition over the 0.5--11.5 GeV/$c$ range has recently been presented in Ref.~\cite{vanDijk:2025ggb}. For the momenta relevant to this work (0.5, 1, 2, 3 GeV/$c$), the corresponding electron fractions are about 97.3\%, 86.7\%, 49.1\%, and 29.0\%.
The beam intensity was adjusted via a collimator and could be varied between $10^3$$\sim$$10^7$ particles per spill, with momentum spread values ranging from 0.6\% to 15\% depending on the momentum setting.

\subsection{Detector Setup}

The overall detector setup is shown in Fig.~\ref{fig:setup}.
From upstream to downstream, threshold Cherenkov counters and scintillating counters for beam monitoring were located in the experimental hall, followed by two Delay Wire Chambers (DWCs)~\cite{Spanggaard:1998}.
A trigger scintillator system was installed between the DWCs, and the Pb/SciFi calorimeter was positioned at the most downstream location.

\begin{figure}[hb]
    \centering
    \includegraphics[width=1.0\linewidth]{Figure1.jpg}
    \caption{Experimental setup showing the delay wire chambers, trigger counters, and Pb/SciFi calorimeter, with scintillator counters and Cherenkov counters placed along the beam line.}
    \label{fig:setup}
\end{figure}

The trigger system consisted of two plastic scintillators (EPIC-crystal EPS100) with dimensions of $\rm 1 \times 0.5 \times 10~cm^3$, arranged in a crossed geometry to define an effective overlap area of $\rm 1 \times 1~cm^2$.
Each scintillator was coupled to a Hamamatsu H3168 photomultiplier tube (PMT).
In the CERN PS T10 beam line, threshold Cherenkov gas counters installed in the East Area were used for particle identification on an event-by-event basis~\cite{Bernhard:2021jeb}.
By choosing the gas and adjusting pressure, the Cherenkov threshold can be tuned to separate particle species in the mixed secondary beam.
In this experiment, the Cherenkov counters were operated to identify electrons and suppress contamination from heavier particles.
The DWC is a multi-wire proportional detector widely used in CERN test-beam experiments for charged-particle tracking.
It provides two-dimensional position measurements with a spatial resolution of about 0.2~mm.
Two DWCs (DWC1 and DWC2) were provided by the facility and installed upstream of the calorimeter to reconstruct the particle trajectory.
The position and angular information obtained from the DWCs were used to reject events affected by scattered background particles in the analysis.

\subsection{Pb/SciFi Calorimeter Prototype}
\begin{figure}[h]
    \centering
    \includegraphics[width=0.85\linewidth]{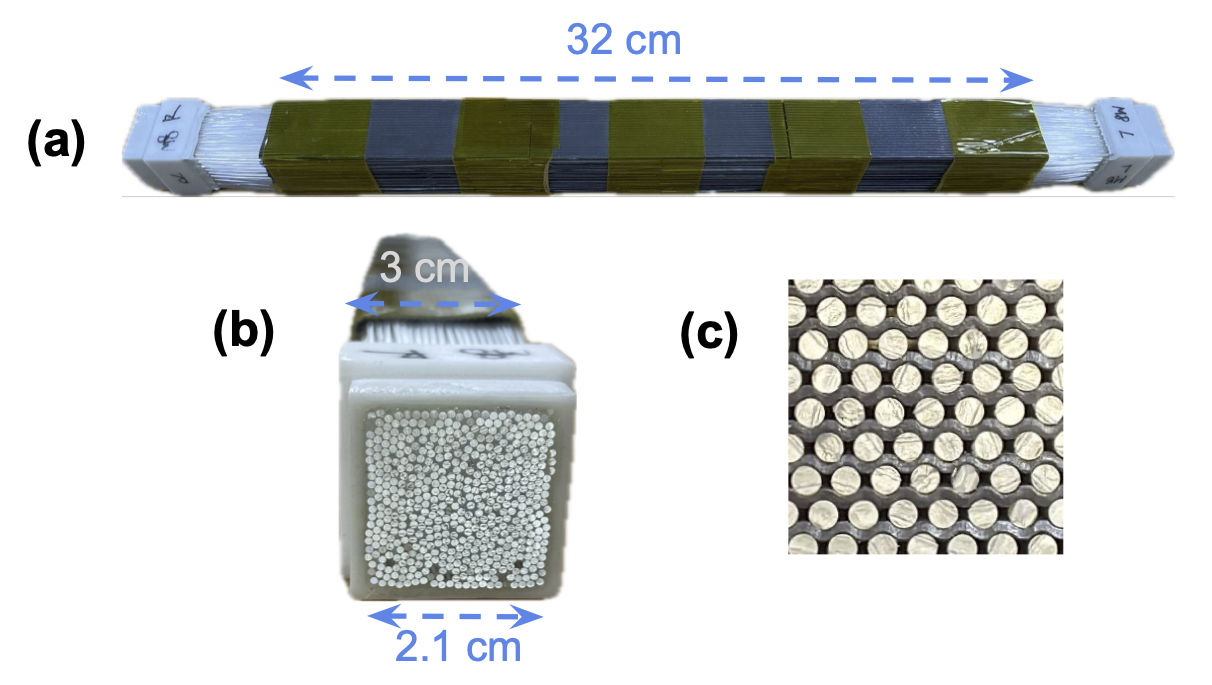}
    \caption{(a) A Pb/SciFi calorimeter unit module with a dimension of $\rm 32\times3\times3\mathrm{\ cm}^3$. (b) Bundled fiber edge using 3D printed case with an active area of $2.1\times2.1\mathrm{\ cm}^2$. (c) The cross section of a unit module.}
    \label{fig:module}
\end{figure}

For the experiment, a set of Pb/SciFi unit modules was produced. Each unit module has dimensions of $32\times3\times3\mathrm{\ cm}^3$. During production, the swaged 0.5-mm-thick lead sheets and scintillating fibers (Kuraray SCSF-78) were stacked alternately  as illustrated in Fig.~\ref{fig:module}, with the lead sheets oriented perpendicular to the beam direction. The material composition is 40:43:17 for Pb:SciFi:Air, which is slightly different from that of the BIC design (37:43:20 for Pb:SciFi:Glue). In contrast to the GlueX BCAL, the Baby BCAL prototype reported in Ref.~\cite{Klest:2025pza}, and the BIC design, the scintillating fibers in the present prototype were not glued into the grooves of the lead sheets. The radiation length ($X_0$) of a unit module is 1.38 cm. At both ends, the fibers were bundled into a $\rm 2.1\times2.1\ cm^2$ area, extending about 5 cm from the edges of the lead sheets as shown in Fig.~\ref{fig:module}. The bundled fibers were then cured with optical glue (Eljen EJ-500) and polished. Both ends of the fiber bundles were coupled to glass PMTs (Hamamatsu R11265-100) using 3D-printed holders and optical cookies were applied to ensure optical coupling.

\begin{figure}[h]
    \centering
    \includegraphics[width=0.85\linewidth]{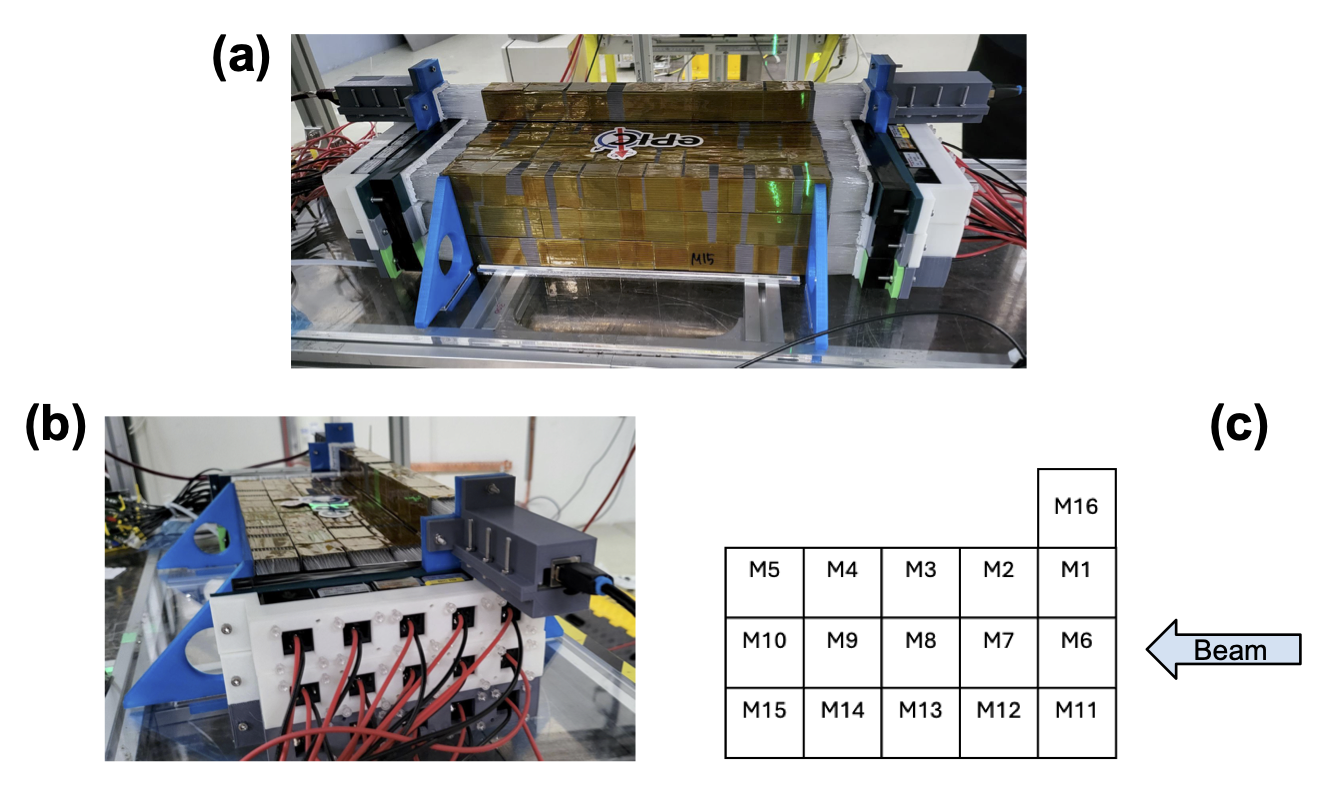}
    \caption{(a) Pb/SciFi calorimeter setup. 3$\times$5 array of unit modules was installed with a total dimension of $\rm 32(w)\times9(h)\times15(d)\mathrm{\ cm}^3$. (b) Side view of the modules. A unit module equipped with SiPMs was placed on top of the 3$\times$5 stack, but it is not included in this study. (c) Layout of the Pb/SciFi modules with respect to the beam direction.}
    \label{fig:assembly}
\end{figure}

The calorimeter setup for the experiment was constructed by stacking unit modules in a 3$\times$5 configuration, where 3 corresponds to the number of transverse rows and 5 corresponds to the number of modules along the beam direction, as shown in Fig.~\ref{fig:assembly}. The total dimension of the assembled calorimeter was $\rm 32(w)\times9(h)\times15(d)\mathrm{\ cm}^3$. The total depth (15 cm) corresponds to 10.9 $X_0$. 

It should be noted that the Pb/SciFi prototype used in this study differs in several respects from the final BIC design. As indicated above, it has a smaller longitudinal depth and limited transverse coverage. The present prototype employed direct readout with glass PMTs in this experiment, whereas the BIC design adopts SiPM readout with light guides, resulting in a different optical coupling and light-sharing configuration. In addition, the unit-module geometry introduces response non-uniformities that may affect the measured performance. The results should therefore be understood in the context of the present prototype configuration.

\subsection{Data Acquisition}

The data acquisition (DAQ) system, produced by NOTICE Korea, consists of a Trigger Control Board (TCB) and digitizer modules using Domino Ring Sampler 4 (DRS4) chip~\cite{Ritt:2004gm,Cho:2025tbq}. Each DAQ module provides 32 input channels and digitizes analog signals with a dynamic range of 1~V mapped to a 12-bit ADC, together with 10-bit time sampling at 5~GHz, corresponding to a sampling interval of 0.2~ns for this measurement. Two DAQ modules were employed: the first allocated two channels for trigger signals and 30 channels for the Pb/SciFi calorimeter, while the second accommodated two channels for the Cherenkov counters and four channels for the horizontal and vertical readout of the two DWCs.

\begin{figure}[h]
    \centering
    \begin{tabular}{cc}
\includegraphics[width=0.7\textwidth ]{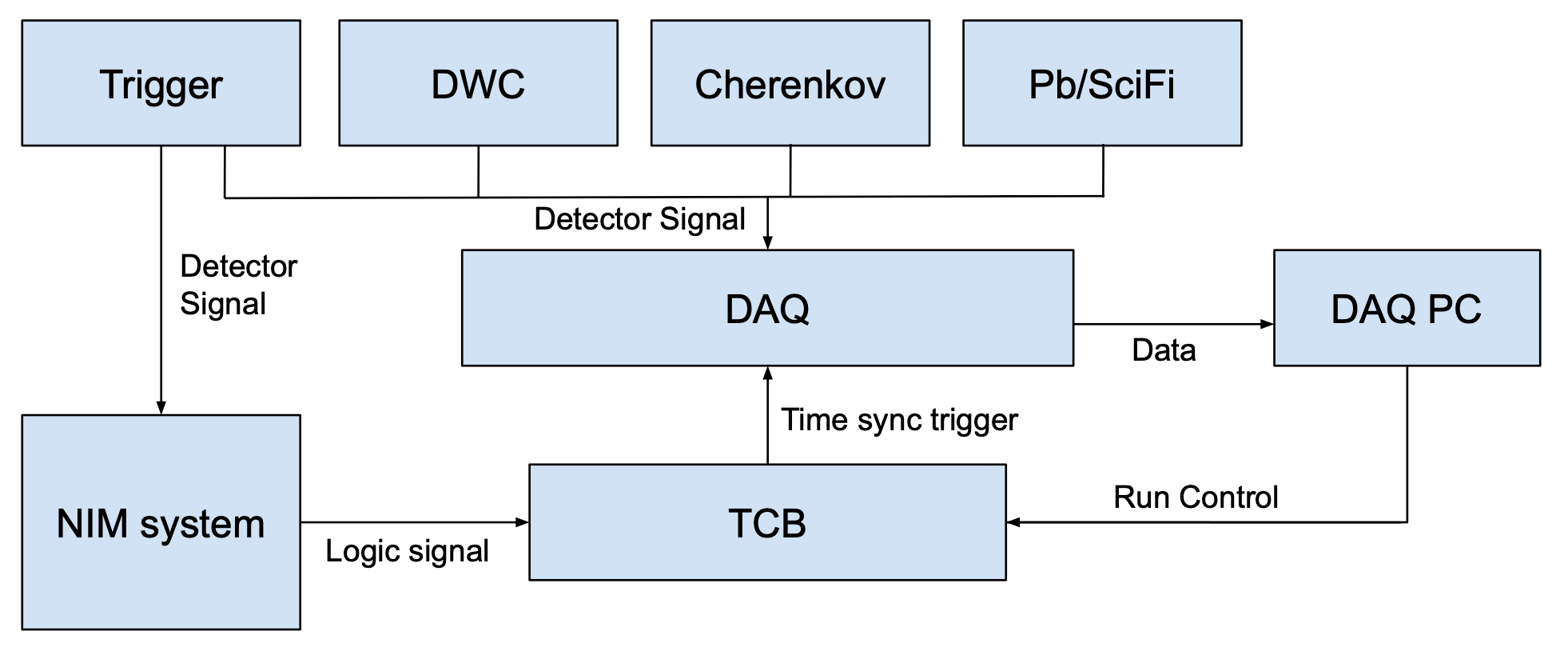}& \includegraphics[width=0.2\textwidth ]{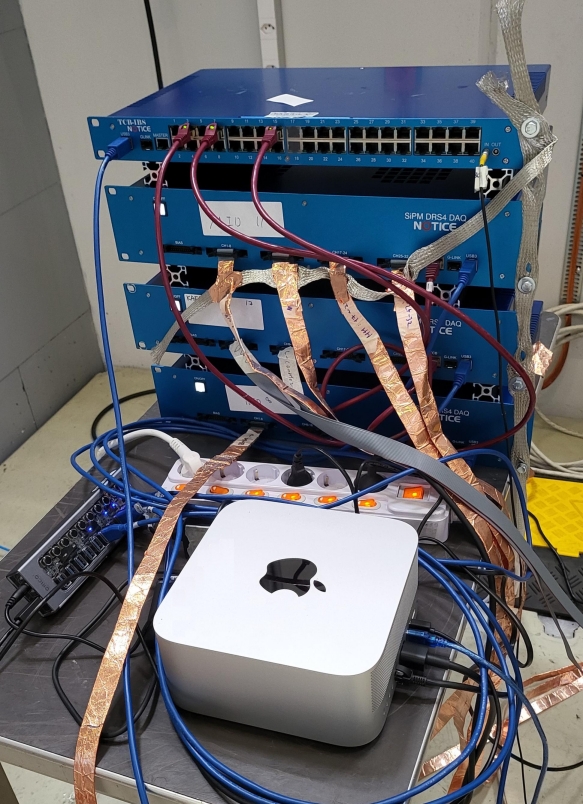}\\ 
\end{tabular}
    \caption{DAQ scheme in the beam test and DAQ system with PC.}
    \label{fig:scheme}
\end{figure}

Figure~\ref{fig:scheme} shows the DAQ scheme. Raw signals from the two trigger detectors were sent to a Nuclear Instrumentation Module (NIM) chain. When both trigger detectors were fired, a coincidence signal was generated after discrimination and logic processing. This logic signal was delivered to the TCB, which issued a synchronized trigger to all DAQ modules, enabling simultaneous digitization across all channels. The TCB and DAQ modules were interconnected via RJ-45 cables for communication, and data were transferred from DAQ modules to the DAQ PC through USB-C connections. The detector signals were interfaced to the DAQ modules using adapter boards that convert LEMO connectors to flat-cable format.

\subsection{Signal Processing and Event Selection}
\begin{figure}[h!]
    \centering
    \includegraphics[width=0.9\linewidth]{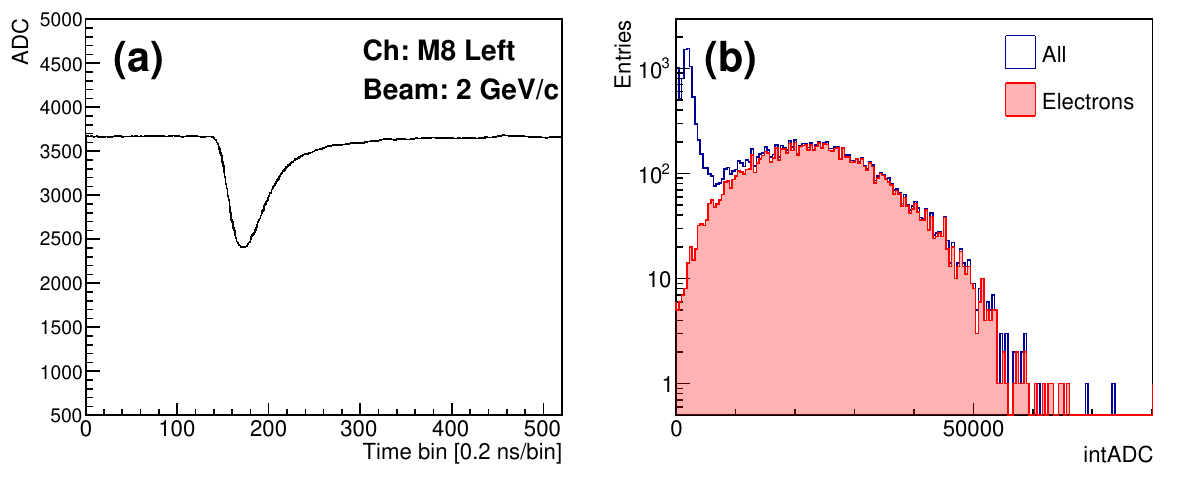}
    \caption{(a) A signal waveform in an event for a channel (M8 in Fig.~\ref{fig:assembly}) at 2 GeV/$c$ beam. The pedestal is calculated by averaging ADC values from first to 120-th time bin. The integrated ADC is obtained from sum of ADC values from 121 to 350-th time bin after subtracting pedestal. (b) Integrated ADC (intADC) distribution of the same channel. The intADC distribution after electron identification using Cherenkov counters is shown in red shaded histogram.}
    \label{fig:pulse}
\end{figure}

The digitized pulse waveform recorded by the DAQ PC for a single event in the readout channel at one end of module M8 in Fig.~\ref{fig:assembly} is shown in the left panel of Fig.~\ref{fig:pulse}.
The pedestal is determined as an averaged ADC value from first to 120-th time bin range. After subtracting the pedestal, the integrated ADC (intADC) is determined by summing ADC values within 121 and 350-th time bin. The pedestal and integration windows were chosen empirically from the observed pulse shape so as to include the full signal. The distribution of the integrated ADC is shown on the right side of Fig.~\ref{fig:pulse}. The distribution of electron identified by Cherenkov counters is shown as red shaded histogram. 

\begin{figure}[h!]
    \centering
    \includegraphics[width=0.95\linewidth]{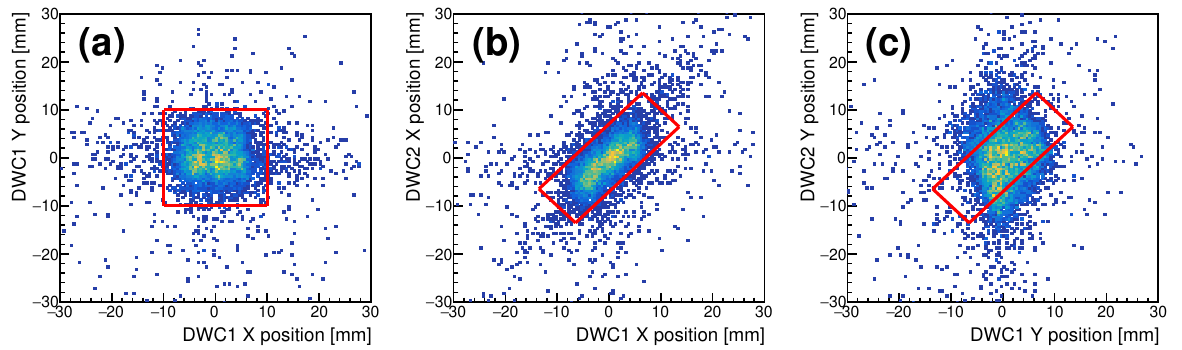}
    \caption{
    (a) Horizontal (X) and vertical (Y) position of the beam particle (2 GeV/$c$) at the upstream DWC (DWC1). A loose cut $\pm 10$~mm is applied in both DWC1 and downstream DWC (DWC2). (b) Correlation of horizontal position between DWC1 and DWC2. Events within a distance of $\pm5$~mm from the diagonal line are selected. (c) Correlation of vertical position between DWC1 and DWC2. Events within a distance of $\pm5$~mm from the diagonal line are selected.}
    \label{fig:DWC}
\end{figure}

After the pulse processing, additional event-selection criteria were applied using the DWC information to reject scattered background particles. To reject background, the beam position was monitored using two DWCs located upstream (DWC1) and downstream (DWC2).
The left panel of Fig.~\ref{fig:DWC} shows the horizontal (X) and vertical (Y) positions of beam particles measured by DWC1 from 2~GeV/$c$ electrons identified by the Cherenkov counters.
A loose selection of $\pm10$~mm was applied as shown by the box in the leftmost panel.
The correlations for the horizontal and vertical positions between the positions at DWC1 and DWC2 are shown in Fig.~\ref{fig:DWC}. The beam conditions were similar across the studied energies. A small shift of the beam center at 0.5~GeV/$c$ (about 3~mm within the 20$\times$20~mm$^2$ window) was checked by varying the position cut and was found to have no effect on the energy resolution.
In the upstream-downstream position correlation, a selection of a distance within $\pm5$~mm from the diagonal line was applied for the horizontal and vertical correlation, taking into account the available statistics and energy resolution. Given the 1.28~m separation between DWC1 and DWC2, this requirement corresponds to a maximum incident angle of about 4 mrad in each projection, ensuring nearly normal incidence at the calorimeter. 

\section{Calibration}
\label{sec:calib}

\subsection{Beam Alignment and Channel Equalization}
Prior to channel equalization, the calorimeter was aligned with respect to the beam by horizontally and vertically scanning the Pb/SciFi modules with a motorized table. The array boundaries were identified from the decrease in gain observed at the module edges, and the beam center was then adjusted to the geometrical center of the 3$\times$5 module array.

The gains of the PMTs coupled to the Pb/SciFi modules were equalized by adjusting their bias voltages using a 2~GeV/$c$ electron beam. A \textsc{Geant4} simulation (v10.05.p01) incorporating the prototype geometry and material composition was used to estimate the expected mean energy deposited in the full volume of each Pb/SciFi module~\cite{GEANT4:2002zbu}. For a 2~GeV/$c$ electron beam, the predicted mean total energy deposits in the five beam-line modules along the beam direction were 129.3, 414.5, 443.3, 299.5, and 162.8~MeV from upstream to downstream.
During the equalization procedure, the beam was centered on each transverse row (M1--M5, M6--M10, and M11--M15) in turn, and the PMT voltages were iteratively adjusted to equalize the left and right readout responses of each module while maintaining a consistent ratio between the integrated ADC counts and the simulated energy deposits for the five modules. The initial high-voltage setting was approximately 600~V, and the final settings were chosen to avoid ADC saturation under the highest signal conditions encountered during the measurements. The resulting high-voltage settings should therefore be understood as those adopted for the present beam-test conditions, rather than as a fully optimized operating point for the detector response.

\subsection{Calibration Procedure}

The calibration of the Pb/SciFi calorimeter channels was performed in offline analysis following the channel equalization procedure to establish a consistent energy scale. Several calibration strategies were tested using the same recorded data samples, and the method providing the best energy resolution was adopted for the final results. In all calibration schemes, a calibration constant for each channel was derived by matching the measured integrated ADC (intADC) to the corresponding mean energy deposit predicted by the same \textsc{Geant4} simulation.

The calibration method adopted for the final results was based on simulations including both electrons and pions. In this approach, pions were included according to the particle fractions reported in Ref.~\cite{vanDijk:2025ggb}, and the resulting simulated energy deposits were compared to the measured intADC. The beam was directed onto the centers of the three transverse rows (M1--M5, M6--M10, and M11--M15) by adjusting the beam position. The mean intADC for each channel was then matched to the corresponding energy deposit. This method provided the best energy resolution among the tested methods.

For comparison, an electron-based calibration method was studied, motivated by the early stage of the run when the Cherenkov counters were not available. In this case, the event selection was based on an electron-like threshold in the intADC distributions (e.g., intADC $>$ 8000 in Fig.~\ref{fig:pulse}). After the Cherenkov counters became operational, an additional study was performed with the beam directed only onto the central transverse row (M6--M10). Calibration constants were also derived for the top and bottom transverse rows using a simulation in which the beam traversed the central transverse row. However, these approaches showed worse energy resolution compared to the adopted method, with an approximately 28\% larger relative width at 2 GeV/$c$.

After determining the calibration constants for the 30 Pb/SciFi calorimeter channels, the reconstructed energy of each module was obtained by summing the calibrated energies from the left and right readout channels, each corresponding to half of the module energy under the equal-sharing assumption. The total reconstructed energy was then calculated by summing the module energies over all modules. An additional global scale factor was applied as a residual overall normalization factor to account for the mismatch between the summed reconstructed energy and the total energy deposit in the simulation. After calibration, data for the energy resolution study were collected using a collimator configuration optimized to minimize the beam momentum spread.

\section{Data Analysis and Results}
\label{sec:results} 
\subsection{Energy Resolution and Linearity}
\begin{figure}[!b]
    \centering  
\includegraphics[width=0.7\linewidth]{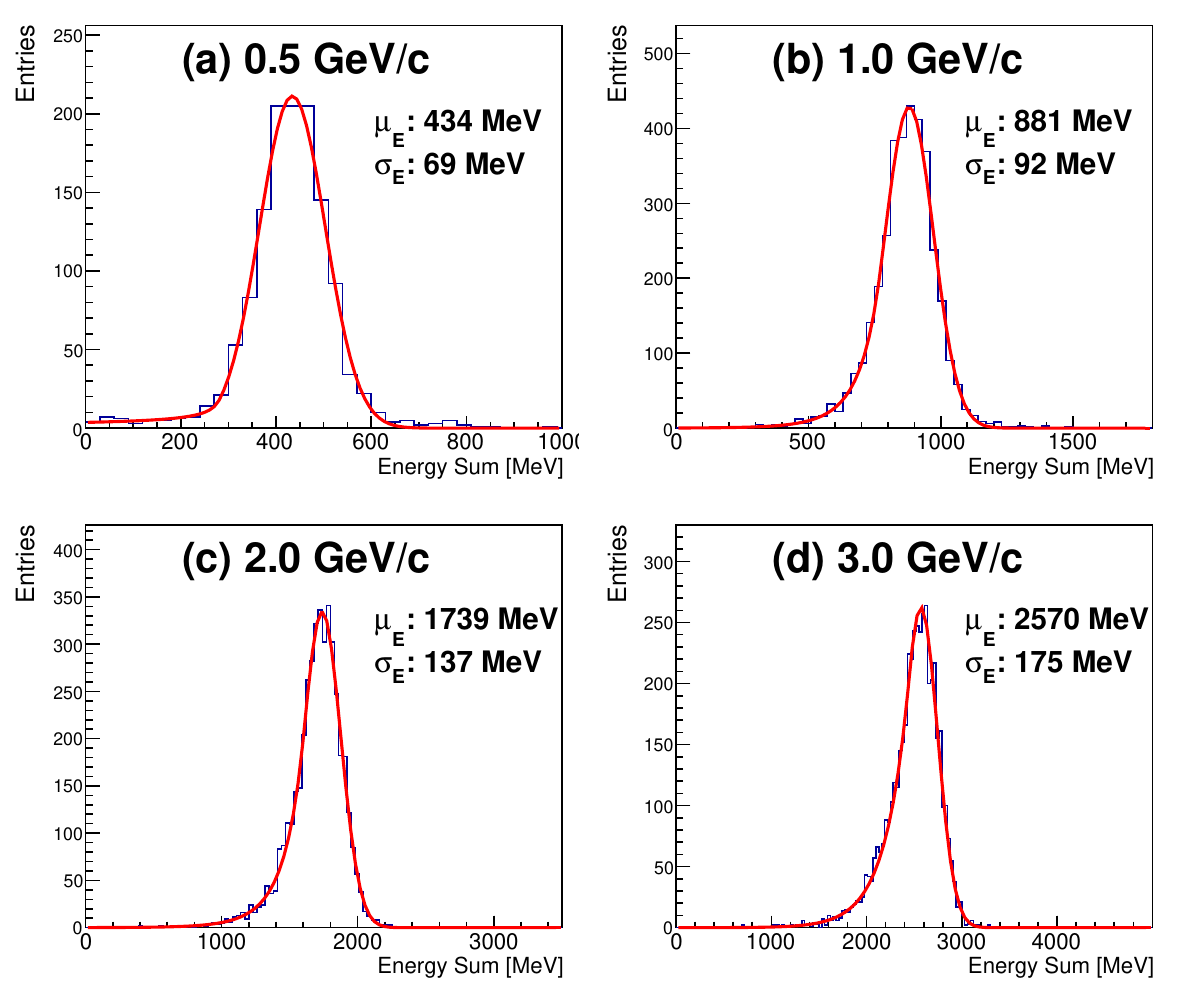}
    \caption{Reconstructed energy for electrons at 0.5, 1, 2, and 3 GeV/$c$. The Gaussian width $\sigma_{E}$ and measured energy $\mu_{E}$ are obtained from a Crystal Ball function, consisting of a Gaussian core and a power-law tail.}
    \label{fig:Edepo}
\end{figure}
After electron selection using Cherenkov counters and event selection using DWCs, the reconstructed energies in the Pb/SciFi calorimeter for electron beams with momenta of 0.5, 1, 2, and 3~GeV/$c$ are shown in Fig.~\ref{fig:Edepo}. 
A low-energy tail becomes more pronounced at higher beam energies, particularly at 3~GeV/$c$, compared to lower energies, which is likely driven by increased longitudinal leakage in the finite-depth prototype. 
In contrast, the 0.5~GeV/$c$ distribution shows a broader low-energy-side excess, which may reflect the smaller signal size and the correspondingly larger relative impact of pedestal subtraction, calibration, and detector non-uniformities.
The distributions are fitted with a Crystal Ball function, consisting of a Gaussian core and a power-law tail, to extract the peak energy and width.
The Gaussian width parameter $\sigma_{E}$ and measured energy $\mu_{E}$ are used to determine the energy resolution.

\begin{figure}[h]
    \centering
    \includegraphics[width=0.9\linewidth]{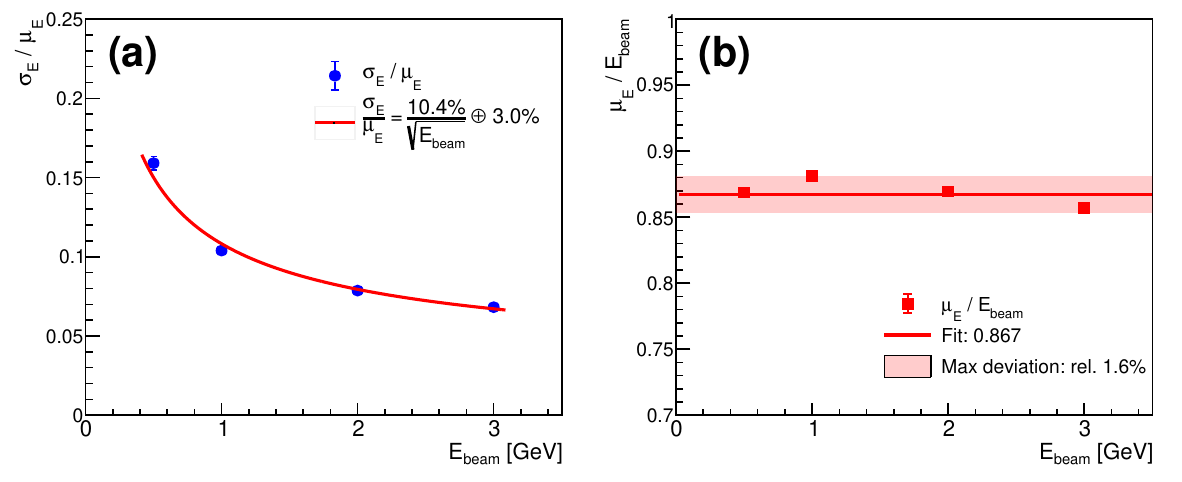}
    \caption{(a) Energy resolution as a function of beam energy. (b) Ratio of the measured energy ($\mu_E$) to the beam energy ($E_{\mathrm{beam}}$) as a function of beam energy. A constant fit gives 0.867, with deviations within 1.6\% over the investigated energy range. The error bars are smaller than the marker size.}
    \label{fig:Eres}
\end{figure}

The energy resolution obtained for electron beams at 0.5, 1, 2, and 3~GeV/$c$ is shown in the left panel of Fig.~\ref{fig:Eres}. 
The dependence of the energy resolution on beam energy is parameterized as
\[
\frac{\sigma_{E}}{\mu_{E}} = \frac{a}{\sqrt{E_{\rm beam}}} \oplus b,
\]
where $\oplus$ indicates addition in quadrature, $E_{\rm beam}$ is beam energy, and $a$ and $b$ represent the stochastic and constant terms, respectively. The fit yields a stochastic term of 10.4\% and a constant term of 3.0\% with $\rm \chi^2/ndf = 9.00/2$. No correction for the beam momentum spread was applied in the present analysis.
When an additional noise term of $\mathrm{c}/E_{\rm beam}$ is considered, it results in 7.5\% of stochastic term, 5.0\% of constant term, and 5.3\% of noise term with $\rm \chi^2/ndf=0.61/1$.
In the present energy range up to 3~GeV/$c$, the stochastic contribution remains dominant, making the constant term difficult to constrain independently. In addition, the fitted resolution components are already correlated in both fits. Together with the limited number of data points, this makes the individual resolution components difficult to constrain in the present measurement.

The measured energy resolution is sensitive to the calibration and equalization conditions of the present beam test. In particular, the initial calibration was constrained by the absence of Cherenkov-based electron tagging, and the equalization was performed under mixed-beam conditions. In addition, the study based on the central transverse row suggests that the voltage settings and calibration conditions used in the present beam test were not fully optimized for modules with only small induced energy deposits, even when Cherenkov-based electron selection was available. An improved energy resolution is therefore expected with refined calibration methods and more controlled electron-beam conditions.

The right panel of Fig.~\ref{fig:Eres} shows the ratio of the measured energy ($\mu_{E}$) to the beam energy ($E_{\rm beam}$), which is used here to characterize the linearity of the calorimeter response over the investigated energy range. 
A constant fit gives 0.867, with deviations within 1.6\% relative to the fit result.
In a finite-depth prototype, this ratio is expected to decrease at higher beam energies because of increasing leakage. The slight deviation of the 0.5~GeV/$c$ point from a simple monotonic leakage trend is likely related to the calibration and equalization conditions of the present beam test, which were optimized using 2~GeV/$c$ data, together with the larger relative impact of small-signal effects at low energy.

\subsection{Longitudinal shower development}

\begin{figure}[h]
    \centering
    \includegraphics[width=1.0\linewidth]{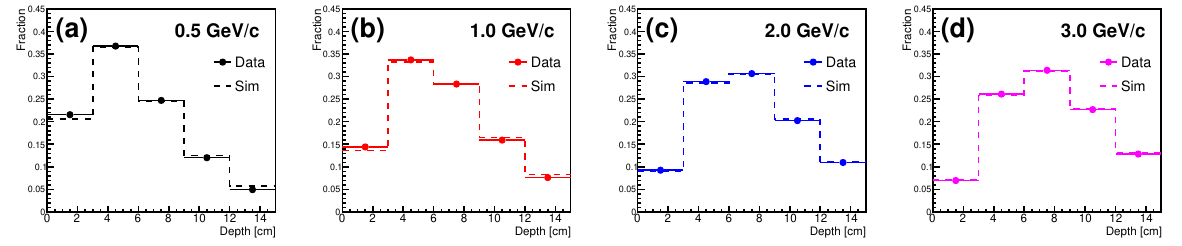}
    \caption{Normalized energy fractions in five unit modules (M6, M7, M8, M9, and M10 in Fig.~\ref{fig:assembly}) for 0.5, 1, 2, and 3 GeV/$c$ electrons, with the beam aligned to pass through the centers of the modules, propagating from upstream (left) to downstream (right).}
    \label{fig:Longitudinal}
\end{figure}

The longitudinal shower development in the Pb/SciFi structure is studied using the fractional energy deposition among modules along the beam direction.
For this analysis, only the modules in the central transverse row (M6--M10) are considered, where the beam is centered on the calorimeter and traverses from left (M6) to right (M10) in Fig.~\ref{fig:Longitudinal}. Although the shower extends beyond the central row, this row is directly traversed by the beam and provides the most reliable calibration in the present data set.
For both data and simulation, the observable is defined as the mean energy deposit in each module normalized to the sum of the mean energy deposits over modules M6--M10, yielding the fractional energy deposition along the beam direction. A normalized representation was adopted to emphasize the longitudinal shower shape, since a direct comparison of the absolute module-by-module energy deposits was found to be more sensitive to residual calibration effects, particularly at 0.5~GeV/$c$.
As expected from the simulation, the position of the maximum energy deposit shifts deeper into the calorimeter with increasing beam energy. The remaining differences between data and simulation are likely influenced by the calibration procedure. The calibration was derived using a mixed electron--pion sample, whereas the longitudinal-shower comparison uses data taken under slightly different beam and run conditions. In addition, since the calibration was anchored at 2~GeV/$c$, small residual mismatches in the normalized longitudinal response can remain at other beam energies.

\subsection{Timing Resolution}

\begin{figure}[h!]
    \centering
    \includegraphics[width=0.8\linewidth]{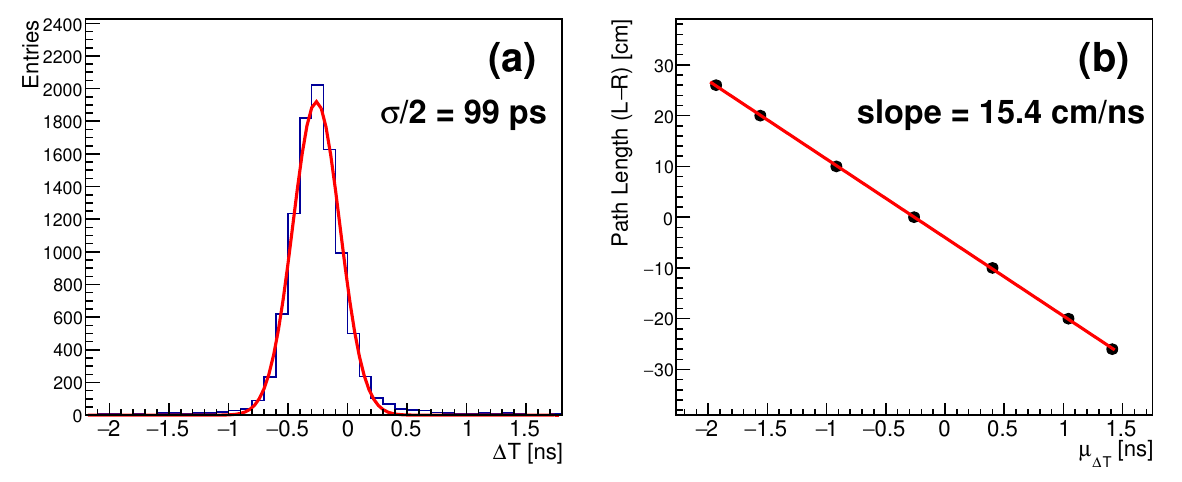}
    \caption{An example of time-difference measurement ($\Delta T = T_L - T_R$) between left and right readout channels using 2 GeV/$c$ electrons. (a) $\Delta T$ distribution used to extract the time resolution when the beam position is at the center of the module, yielding $\sigma_{\Delta T}/2 = 99$ ps. (b) Path-length difference between the left and right readout as a function of the mean time difference ($\mu_{\Delta T}$) for various horizontal beam positions ($\pm13, \pm10, \pm5, 0$ cm), used to determine an effective horizontal propagation speed of 15.4 cm/ns. A constant offset in $\mu_{\Delta T}$ originates from fixed DAQ timing.}
    \label{fig:time}
\end{figure}

The timing resolution and effective photon-propagation speed of the Pb/SciFi modules were evaluated using the timing difference between signals read out from the left and right sides of each module, using 2~GeV/$c$ electrons.
The left panel of Fig.~\ref{fig:time} shows the time difference between the left and right readout channels ($\Delta T = T_L - T_R$) where the beam traverses the center of the modules. 
The timing of each pulse is defined as the moment when the waveform reaches 30\% of its peak amplitude, which was found to provide better timing resolution than other constant-fraction values. 
The timing resolution of the module hit time is obtained from a Gaussian fit to the $\Delta T$ distribution as $\sigma_{\Delta T}/2$~\cite{Leverington:2008zz}.  An example of the $\Delta T$ distribution for one module (M9) is shown in the left panel of Fig.~\ref{fig:time}. For the studied modules (M6, M7, M8, M9), resulting values of $\sigma_{\Delta T}/2$ range from 99 to 135~ps. 
The non-zero offset of $\Delta T$ reflects fixed DAQ timing offsets, and remained stable throughout the measurement.

The timing information also enables a study of the effective photon-propagation speed in the scintillating fibers. 
Photons travel through the scintillating fibers at a nominal speed of $c/n \approx 18.8$~cm/ns for a polystyrene core with refractive index $n=1.59$, but internal reflections reduce the effective horizontal propagation speed. 
To quantify this, data were taken at several horizontal beam positions ($\pm13,\,\pm10,\, \pm5,\,0$ cm), corresponding to left-right path-length differences of $\pm26,\pm20,\pm10,0$~cm. 
For each position, the mean time difference $\mu_{\Delta T}$ was extracted from a Gaussian fit. 
The correlation between path-length difference and $\mu_{\Delta T}$ is shown in the right panel of Fig.~\ref{fig:time}. 
A linear trend is observed, and linear fits to the seven points yields an effective photon-propagation speeds ranging from 15.3 to 15.7~cm/ns. 
Only uncertainties on $\mu_{\Delta T}$ are included in the plot; the horizontal beam size, approximately $1\times1$~cm$^{2}$ for the triggered events, was not propagated into the uncertainty. Combining the effective speed with the timing resolution gives an estimated position resolution of approximately 1.5~cm for this module under the conditions of this experiment.

\section{Summary}
\label{sec:summary}
The first beam test of a Pb/SciFi calorimeter prototype produced in Korea for the Barrel Imaging Calorimeter (BIC) of the EIC was carried out at the CERN PS T10 beam line using low-energy electron beams.
This measurement represents an initial experimental validation of the Pb/SciFi unit-module concept under realistic beam conditions.
The detector response was characterized in terms of energy deposition, longitudinal shower development, and timing performance. Several calibration strategies were also explored.
In the 0.5--3~GeV/$c$ electron momentum range, the prototype shows good energy resolution for a small-scale setup.
The measured performance should be interpreted in the context of the present prototype configuration, which differs from the final BIC design in depth, transverse coverage, and readout scheme. Future work will focus on improved equalization and calibration under more controlled electron-beam conditions, as well as on the construction and beam test of a larger-scale prototype with geometry and readout closer to the final BIC design.
The results provide important guidance for future beam tests and for the development of larger-scale prototypes, including imaging layers equipped with AstroPix silicon pixels and Pb/SciFi layers, toward the realization of a full-scale BIC.

\section*{Acknowledgments}
This work was supported by the National Research Foundation of Korea (NRF) grants funded by the Korean government (MSIT) (Grant Nos. RS-2024-00408265, 2020R1A2C3013540, RS-2025-00514606, RS-2024-00355188, RS-2021-NR062024, RS-2023-00280845, RS-2018-NR031074, RS-2022-NR070380, and RS-2023-00279977).
This work also received funding from the European Union's Horizon Europe Research and Innovation Programme under Grant Agreement No. 101057511 (EURO-LABS). 
The material is based upon work partially supported by the U.S. Department of Energy, Office of Science, Office of Nuclear Physics and Laboratory Directed Research and Development funding from Argonne National Laboratory, provided by the Director, Office of Science, of the U.S. Department of Energy under Contract No. DE-AC02-06CH11357.
This work was also supported by the Natural Sciences and Engineering Research Council of Canada under Grant No. SAPPJ-2025-00040.
The authors would like to thank the CERN PS team for their support.

\end{document}